\documentclass[utf8]{frontiersinFPHY_FAMS}
\usepackage{hyperref}
\usepackage[numbers]{natbib}
\usepackage{url,hyperref,lineno,microtype,subcaption}
\usepackage[onehalfspacing]{setspace}
\usepackage{bm}
\usepackage{parskip}[=v1]

\def\firstAuthorLast{Gill \& Lambare} %use et al only if is more than 1 author
\def\Authors{Richard D. Gill\,$^{1,*}$, Justo Pastor Lambare,$^{2}$}

\def\Address{$^{1}$Mathematical Institute, Leiden University, Netherlands\\
$^{2}$Facultad de Ciencias Exactas y Naturales, Ruta Mcal. J. F. Estigarribia, Km 11 Campus de la UNA, San Lorenzo, Paraguay}

\def\corrAuthor{Richard Gill, Mathematical Institute, Leiden University, Netherlands}

\def\corrEmail{gill@math.leidenuniv.nl}

\begin{document}
\onecolumn
\firstpage{1}

\title[Commentary: Is the moon there if nobody looks]{Commentary: Is the moon there if nobody looks -- Bell inequalities and physical reality }

\author[\firstAuthorLast ]{\Authors} %This field will be automatically populated
\address{} %This field will be automatically populated
\correspondance{} %This field will be automatically populated

\extraAuth{}% If there are more than 1 corresponding author, comment this line and uncomment the next one.
%\extraAuth{corresponding Author2 \\ Laboratory X2, Institute X2, Department X2, Organization X2, Street X2, City X2 , State XX2 (only USA, Canada and Australia), Zip Code2, X2 Country X2, email2@uni2.edu}

\maketitle

\section{Introduction}

Marian Kupczynski (``MK'') is the author of a controversial paper published (2020) in the journal {\em Frontiers in Physics} \cite{MK1}.
The work is built around a mathematical claim by MK which is actually false, and MK's reasoning around this claimed assertion is also false.
The proof of that is the main content of our present ``Comment''.
It is merely a mathematical counter-example to a mathematical claim in \cite{MK1}.

\section{Bell's inequalities do hold for Kupzcynski's model}
Kupczynki \cite{MK1} incorporates hidden variables, standing for random disturbances arising in the measurement apparatus and dependent on the local measurement setting, as follows. Consider an experiment in which Alice and Bob's settings will be $x$ and $y$. To begin with, hidden variables $(\lambda_1, \lambda_2)$ with some arbitrary joint probability mass function $p(\lambda_1, \lambda_2)$, not depending on the local settings $x$ and $y$ chosen by the experimenters, are transmitted from the source to the two measurement stations. At Alice's station and Bob's station, independently of one another, and independently of $(\lambda_1, \lambda_2)$, local hidden variables $\lambda_x$ and $\lambda_y$ are created with probability mass functions $p_x(\lambda_x)$ and $p_y(\lambda_y)$. The measurement outcome on Alice's side is then $A_x(\lambda_1, \lambda_x)$, and similarly on Bob's side, $B_y(\lambda_1, \lambda_y)$. The functions $A_x$ and $B_y$ depend in any way whatever on $x$ and $y$ respectively; even the domains of these functions can vary. The sets of possible outcomes of $\lambda_x$ and $\lambda_y$ may depend on $x$ and $y$ respectively. Now repeat this story with, instead of $x, y$, settings $x, y'$, then $x', y$, then $x', y'$. In this way, Kupczynski has defined the four expectation values $E(A_{x} B_{y})$, $E(A_{x'} B_{y})$, $E(A_{x} B_{y'})$, $E(A_{x'} B_{y'})$ of interest, on four ``dedicated'' hidden variable spaces, which he moreover states are ``disjoint''.
Therefore, he is unable to define certain ``counterfactual'' expectations which are used in his proof in the non-contextual case of the Bell-CHSH inequalities. Does this mean that the inequalities need not hold? His argument that they could be violated is based on the huge number of free parameters which his model allows. However, he does not actually specify any particular instantiation of all those parameters which does the job. He does claim that other authors did already do just that.

MK says about his framework: ``counterfactual expectations $E(A_xA_{x'})$, $E(B_yB_{y'})$, $E(A_xA_{x'}B_yB_{y'})$ do not exist and Bell and CHSH inequalities may not be derived''. He hereby refers to the usual CHSH inequalities for the four expectations $E(A_xB_y)$, $E(A_xB_{y'})$, $E(A_{x'}B_y)$, $E(A_{x'}B_{y'})$.
The context is a Bell-type experiment in which Alice chooses between settings $x$ and $x'$, and Bob chooses between settings $y$ and $y'$. MK talks about four different Kolmogorov probability models for the four sub-experiments (one setting choice for Alice and one for Bob). Here are his expressions for the four expectation values of interest, where I have amplified his already long formulas by inserting part of the definition of the four underlying sample spaces $\Lambda_{xy}$,  $\Lambda_{xy'}$,  $\Lambda_{x'y}$,  $\Lambda_{x'y'}$.
$$E(A_{x} B_{y})~=~ \sum\limits_{(\lambda_1, \lambda_2, \lambda_{x}, \lambda_{y}) \in \Lambda_{{x}{y}}} A_{x}(\lambda_1, \lambda_{x})B_{y}(\lambda_2, \lambda_{y})p_{x}(\lambda_{x})p_{y}(\lambda_{y})p(\lambda_1, \lambda_2),$$
$$E(A_{x} B_{y'})~=~ \sum\limits_{(\lambda_1, \lambda_2, \lambda_{x}, \lambda_{y'}) \in \Lambda_{{x}{y'}}} A_{x}(\lambda_1, \lambda_{x})B_{y'}(\lambda_2, \lambda_{y'})p_{x}(\lambda_{x})p_{y'}(\lambda_{y'})p(\lambda_1, \lambda_2),$$
$$E(A_{x'} B_{y})~=~ \sum\limits_{(\lambda_1, \lambda_2, \lambda_{x'}, \lambda_{y}) \in \Lambda_{{x'}{y}}} A_{x}(\lambda_1, \lambda_{x'})B_{y}(\lambda_2, \lambda_{y})p_{x'}(\lambda_{x'})p_{y}(\lambda_{y})p(\lambda_1, \lambda_2),$$
$$E(A_{x'} B_{y'})~=~ \sum\limits_{(\lambda_1, \lambda_2, \lambda_{x'}, \lambda_{y'}) \in \Lambda_{{x'}{y'}}} A_{x'}(\lambda_1, \lambda_{x'})B_{y'}(\lambda_2, \lambda_{y'})p_{x'}(\lambda_{x'})p_{y'}(\lambda_{y'})p(\lambda_1, \lambda_2).$$

These four equations are a complicated way to say the following: with settings $x, y$, hidden variables $(\lambda_1, \lambda_2)$ with some arbitrary joint probability mass function $p(\lambda_1, \lambda_2)$ (independent of $x$ and $y$) are transmitted from the source to the two measurement stations. At Alice's station and Bob's station, independently of one another, local hidden variables $\lambda_x$ and $\lambda_y$ are created with probability mass functions $p_x(\lambda_x)$ and $p_y(\lambda_y)$. The measurement outcome on Alice's side is $A_x(\lambda_1, \lambda_x)$, and similarly on Bob's side, $B_y(\lambda_1, \lambda_y)$. Now repeat this story with,  instead of $x, y$, settings $x, y'$, then $x', y$, then $x', y'$. Kupczynski has defined these expectation values on four ``dedicated'' hidden variable spaces, which he moreover states are ``disjoint''. Therefore, he says, {\em he} is unable to defined certain ``counterfactual'' expectations which {\em his} proof in the non-contextual case used, and hence {\em he} can't derive Bell and CHSH inequalities.

But we can do all that! And we can even use his original proof in the non-contextual case to get those inequalities! Here is just one of many ways.

Take as sample space a set of tuples $\bm\lambda = (\lambda_1, \lambda_2, \lambda_x, \lambda_{x'}, \lambda_y, \lambda_{y'})$. This space is just the Cartesian product of the spaces whose existence Kupczynski already hypothesized. Take as probability mass function on this space the product $p(\lambda_1, \lambda_2) p_x(\lambda_x)p_{x'}(\lambda_{x'})p_y(\lambda_y)p_{y'}(\lambda_{y'})$. Finally, define new measurement functions ${\bf A}(\bm\lambda, x) = A_{x}(\lambda_1, \lambda_{x})$, ${\bf B}(\bm\lambda, y) = B_{y}(\lambda_2, \lambda_{y})$ where $x$ can be replaced by $x'$ and/or $y$ by $y'$. Now compute $E({\bf A}_x {\bf B}_y)$, also with $x$ replaced by $x'$ and/or $y$ by $y'$. It is immediately clear that the four new expectation values of products have exactly the same values as those just exhibited of Kupczynski's. We can now go back to Kupczynski's own earlier traditional derivation of Bell-CHSH. There is no barrier to running through the usual proof since all four expectations of products are defined on the same probability space.

There are more efficient constructions. As one learns in courses on Monte Carlo simulation, one can define a discrete random variable with an arbitrary probability distribution as a function of a single uniformly distributed random variable on the unit interval $[0, 1]$. Thus one could define $\lambda_x$ and $\lambda_x'$ as functions of a single uniformly distributed random variable $U_1$ and of a second argument $x$ or $x'$; similarly define $\lambda_y$ and $\lambda_y'$ as functions of a single, independent, uniformly distributed random variable $U_2$ and of a second argument $y$ or $y'$. We just add to our original $(\lambda_1, \lambda_2)$ two independent random variables $U_1$, $U_2$ and redefine our measurement functions in the obvious way. In this way, we can accomodate any number of setting choices for Alice and Bob without introducing more ``contextual'' randomness into the two measurement functions. This is an important insight. Contextual randomness does not need randomness dependent on the setting. The setting dependence can be passed into the deterministic part of the model.

\section{Discussion}
We have proved that MK's claim that his hidden variables model does not allow Bell's inequalities to be derived is a false mathematical statement.
Furthermore, Mk's paper also contains an obvious contradiction when he inadvertently validates Bell's model stating that \emph{``Although the expectations calculated using the Equations (11--14) and (19--22) have the same values, the two sets of formulas describe different experiments''}. Since both systems of equations \emph{``have the same values''}, MK's hidden variables model also satisfies the Bell inequality.

Kupcynski's blunder arises from a literal interpretation of different equivalent mathematical expressions, one of which has a direct physical meaning, while the other, obtained after correct mathematical transformations, does not.
Curiously, such confusion, which first appeared in 1972 \cite{delapena}, persists to this day \cite{pLam22}.
Refs. \cite{pLam21,pLam21b} explain similar inconsistencies arising from joint probabilities and incompatibility.

Also his criticism of past ``loophople-free'' Bell experiments is unduly harsh and he seems unaware of methodology already in use which minimalises the problem of apparent violation of ``no-signalling''.
A more detailed explanation can be found in a longer and earlier version of this paper available on {\tt arXiv.org} \cite{GL}.

\section*{Conflict of Interest Statement}

The authors declare that the research was conducted in the absence of any commercial or financial relationships that could be construed as a potential conflict of interest.

\section*{Author Contributions}

RDG initiated this project and wrote an initial draft which he shared with JLP.
%It turned out that JLP had already written a critique on another work of Kupczynski together with Rodney Franco \cite{pLam21}.
Discussions led to many changes that resulted on a preprint \cite{GL}.
We decided to summarise our findings in a short ``Comment'' in {\em Frontiers in Physics}. In the meantime, MK published a response \cite{MK2} to our preprint \cite{GL} and a debate ensued on {\tt PubPeer.org}.

\section*{Funding}

The authors received no funding for this research.

\section*{Acknowledgments}

RDG is grateful for numerous discussions with Marian Kupczynski on his work.

\raggedright

\end{document}